\documentclass[onecolumn,showpacs,floatfix]{revtex4}

\newcommand{\figwidth}{0.5\columnwidth}

\usepackage[dvips]{graphicx}
\usepackage{amsmath}
\usepackage{amsfonts}
\usepackage{amssymb}
\usepackage{bm}

\newcommand{\eq}[1]{Eq.(\ref{#1})}
\newcommand{\fig}[1]{Fig.~\ref{#1}}
\newcommand{\tab}[1]{Table~\ref{#1}}

\newcommand{\avg}[1]{ {\langle #1 \rangle} }
\newcommand{\olcite}[1]{Ref.~\onlinecite{#1}}

\newcommand{\nc}{N_{\rm c}}
\newcommand{\zc}{z_{\rm c}}
\newcommand{\zp}{z_{\rm p}}
\newcommand{\vc}{v_{\rm c}}
\newcommand{\sigmac}{\sigma_{\rm c}}
\newcommand{\sigmap}{\sigma_{\rm p}}
\newcommand{\etapr}{\eta_{\rm p}^{\rm r}}
\newcommand{\etaprcr}{\eta_{\rm p,cr}^{\rm r}}
\newcommand{\etaccr}{\eta_{\rm c,cr}}
\newcommand{\pc}{P_L(\nc | \etapr, \zc)}
\newcommand{\etac}{\eta_{\rm c}}
\newcommand{\etacg}{\eta_{\rm c}^{\rm v}}
\newcommand{\etacl}{\eta_{\rm c}^{\rm l}}
\newcommand{\chired}{\chi_{\rm red}}
\newcommand{\Rg}{R_{\rm g}}
\newcommand{\Rc}{R_{\rm c}}
\newcommand{\sigmacp}{\sigma_{\rm cp}}

\begin{document}

\title{Critical behavior in colloid-polymer mixtures: theory and simulation}

\author{F. Lo Verso}
\affiliation{Institut f\"ur Theoretische Physik II, Heinrich-Heine-Universit\"at
D\"usseldorf, Universit\"atsstra{\ss}e 1, D-40225 D\"usseldorf, Germany}

\author{R. L. C. Vink}
\affiliation{Institut f\"ur Theoretische Physik II, Heinrich-Heine-Universit\"at
D\"usseldorf, Universit\"atsstra{\ss}e 1, D-40225 D\"usseldorf, Germany}

\author{D. Pini}
\affiliation{Dipartimento di Fisica, Universit\`a di Milano, Via Celoria 16, 20133 Milano,
Italy}

\author{L. Reatto}
\affiliation{Dipartimento di Fisica, Universit\`a di Milano, Via Celoria 16, 20133 Milano,
Italy}

\date{\today}

\pacs{61.20.Gy, 64.60.Ak, 82.70.Dd, 61.20.Ja, 05.10.Ln,
02.70.-c, 05.70.Jk, 64.60.Fr}

\begin{abstract}

We extensively investigated the critical behavior of mixtures of colloids 
and polymers via the two-component Asakura-Oosawa model and its reduction 
to a one-component colloidal fluid using accurate theoretical and 
simulation techniques. In particular the theoretical approach, 
hierarchical reference theory [\emph{Adv. Phys.} {\bf 44}, 211 (1995)], 
incorporates realistically the effects of long-range fluctuations on phase 
separation giving exponents which differ strongly from their mean-field 
values, and are in good agreement with those of the three-dimensional 
Ising model. 
Computer simulations combined with finite-size scaling
analysis confirm the Ising universality and the accuracy of the theory,
although some discrepancy in the location of the critical point between
one-component and full-mixture description remains. 
To assess the limit of the pair-interaction 
description, we compare one-component and two-component results.

\end{abstract}

\maketitle

E-mail: davide.pini@mi.infm.it   

\section{Introduction}
\label{INTRO}

Mixtures of colloids and polymers are exciting fluid systems. This is 
partly due to their industrial importance, but also to the fundamental 
physical insights they provide \cite{imhof.dhont:1995, poon:2004, 
aarts2004}. An important mechanism governing the phase behavior of 
colloid-polymer mixtures is the depletion effect, which leads to an 
effective attraction between the colloids \cite{ao, 8evans} where the 
polymer fugacity may be regarded as the analogue of inverse temperature. 
In particular, for sufficiently large size ratio $\Rg/\Rc$, where $\Rg$ 
and $\Rc$ are respectively the radius of gyration of the polymer 
and the radius of 
the colloid, a colloid-polymer mixture exhibits stable fluid-fluid phase 
separation and a solid crystal phase. Since colloidal particles can be visualized 
close to single particle resolution using confocal microscopy, exciting 
real-space investigations of fluid criticality are nowadays possible. In 
one such experiment, interface fluctuations were directly visualized 
\cite{aarts2004}. The snapshots given in \olcite{aarts2004} show that the 
interface fluctuations become more pronounced upon approach of the 
critical point, a consequence of the diverging correlation length. In a 
more recent experiment, the critical exponent of the correlation length 
was extracted from real-space data and shown to be compatible with the 
three-dimensional (3D) Ising exponent \cite{royall}. This finding is 
consistent with earlier colloid-polymer experiments 
\cite{chen.payandeh:2000, chen.payandeh:2001}, whose results were 
interpreted in terms of renormalization of the Ising critical exponents 
\cite{fisher}.
Compared to the wealth of results on the overall shape of the phase 
diagram, little theoretical attention 
has been given to the critical behavior of 
colloidal dispersions.
While the universality class is not expected to be changed from the Ising 
one, characteristic of simple fluids, the greater flexibility of 
colloid-colloid interactions with respect to their atomic counterparts may 
allow the study of several nonuniversal aspects of criticality, which are 
difficult to bring forth in atomic fluids.

Inspired in part by the above mentioned experiments, the present paper 
aims to describe fluid criticality in colloid-polymer mixtures 
theoretically. In particular we focused on fluid-fluid phase separation 
and on the critical behavior of the Asakura-Oosawa (AO) model 
\cite{ao,8evans} via numerical simulation and liquid-state theory. On the 
simulation side, we considered the fluid-fluid phase diagram both of the 
AO colloid-polymer mixture, and of the one-component fluid of particles 
interacting via the two-body AO potential for a number of size ratios. The 
critical behavior of the order parameter and of the isothermal 
compressibility were also determined in the binary mixture by finite-size 
scaling techniques.

Despite its simplicity, extracting the critical behavior of the AO model 
by theoretical means is challenging. Characteristic of the mean-field 
approximation used till now is the parabolic shape of the binodal, 
corresponding to $\beta=1/2$ where $\beta$ is the critical exponent of the 
order parameter. Since the AO model belongs to the 3D Ising universality 
class \cite{vink.horbach:2004*1, vink.horbach:2004, vink.horbach.ea:2004}, 
where $\beta \approx 0.326$ \cite{fisher.zinn:1998}, the true binodal is 
flatter. In addition, mean-field approximations appreciably underestimate the 
critical polymer fugacity, such that the location of the critical point is 
not reproduced correctly.

In this paper we studied the critical behavior of the AO pair potential by 
the hierarchical reference theory (HRT) \cite{hrtrev}. Among liquid-state 
theories, the HRT has the peculiar feature of implementing the 
renormalization group machinery, which makes it a reliable tool in the 
study of the critical behavior of fluids, since it takes into account the 
effect of long range fluctuations on phase separation in a 
non-perturbative way. This theory is indeed capable of getting arbitrarily 
close to the fluid-fluid critical point. The present analysis complements 
an earlier investigation \cite{hrt1}, where the stress was more on the 
overall topology of the phase diagram and the stability of fluid-fluid 
phase separation with respect to freezing. Here, we devote our attention 
solely to the fluid critical regime. Notice that we did not make a direct 
comparison with the experimental results of refs. 
\cite{chen.payandeh:2000, chen.payandeh:2001}, since the mixture 
considered there has a polymer-colloid size ratio larger than one. A 
one-component treatment where the polymer degrees of freedom are traced 
out is then untenable in that case.

The main motivations for this paper are the following: first, by comparing 
the fluid-fluid phase diagram of the AO mixture with that of the 
one-component AO fluid with the effective pair interaction, we aim at 
assessing the accuracy of the latter description as the size ratio is 
varied. Second, we want to compare the critical behavior of the AO mixture 
as given by finite-size scaling with the results obtained by applying HRT 
to the AO pair potential. Once reduced quantities are used, so as to make 
allowance for the discrepancy between the critical polymer densities given 
by the one- and the two-component descriptions, we find that the HRT is 
able to reproduce the asymptotic critical power law for the 
compressibility and the order parameter of the AO mixture with good 
accuracy.

The paper is structured as follows: In Sec.~IIA the main features of the 
AO model are recalled. In Sec.~IIB a short overview of the HRT is given. 
Sec.~III is devoted to the illustration of the simulation techniques: 
specifically, Secs.~IIIA and IIIB deal with the cumulant intersection and 
the scaling plots methods for determining the critical temperature and 
amplitudes respectively. In Sec.~IIIC the unbiased scaling algorithm 
mentioned above is described. A biased version of the algorithm, where one 
assumes that the universality class in known, is described in Sec.~IIID. 
In Sec.~IV our HRT and simulation results for several size ratios are 
presented and discussed and our conclusions are drawn.

\section{Introduction to the system and theoretical method of analysis} 
\label{INTRO-sec}

\subsection{The Asakura-Oosawa model}
\label{AO-model}

In this work we considered a binary mixture of colloidal particles and 
non-adsorbing polymers. Neglecting the degrees of freedom of the 
individual solvent molecules and of the polymer monomers, it is useful to 
consider effective potentials between constituents which are pairwise 
additive. In this context a very simple and basic model was proposed by 
Asakura and Oosawa \cite{ao} (AO model) and independently by Vrij 
\cite{8evans}. The colloids are assumed to be hard spheres with radius 
$\sigmac/2$ and the polymers, with radius of gyration $\Rg$, as 
inter-penetrating and non-interacting as regards to their mutual 
interactions. However, the polymers are excluded from the colloids to a 
certain center-of-mass distance $\sigmacp$. Therefore, with respect to 
their interaction with colloidal particles the polymer molecules are 
assumed to behave as hard spheres of radius $\sigmap/2=\Rg$, the diameter 
of the colloid-polymer interaction being $\sigmacp=(\sigmac +\sigmap)/2$. 
The binary AO model is then represented by a binary mixture characterized 
by the following potentials:
\begin{eqnarray}
v_{\rm pp} & = & 0               \nonumber  \\
v_{\rm cp} & = & 
\left\{
\begin{array}{ll}
\infty & r < \sigmacp \\
0 & r>\sigmacp
\end{array}
\right.                     \\  
v_{\rm cc} & = & 
\left\{
\begin{array}{ll}
\infty & r < \sigmac \\
0 & r>\sigmac
\end{array}
\right.                      \nonumber 
\end{eqnarray}
where $r$ is the distance between two particles. The above potentials 
define what we, in this paper, shall call the {\it full mixture} 
description of the AO model, in which the degrees of freedom of {\it both} 
the colloids {\it and} the polymers are explicitly retained.
 
\begin{figure}
\includegraphics[width=\figwidth,clip]{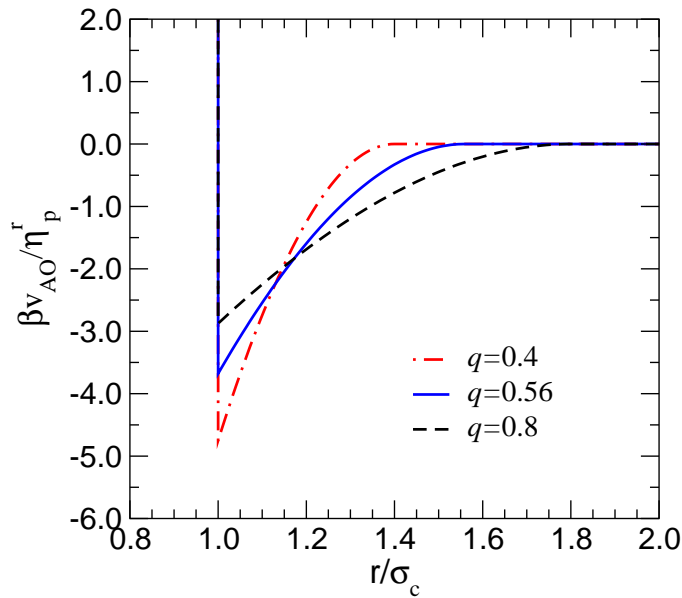}
\caption{\label{AO}Plots of the effective colloid-colloid pair potential,
given by \eq{ao}, as function of the pair separation $r$, for several
colloid-to-polymer size ratios $q$.}
\end{figure}

If the polymer degrees of freedom are traced out, the resulting effective 
(pair) interaction reads as:
\begin{equation}
\label{ao}
\beta v_{\rm AO}(r)=
\left\{
\begin{array}{ll}
\infty & r < \sigmac \\
\displaystyle{-\frac{\pi \sigmap^3 \zp}{6}\frac{(1+q)^{3}}{q^3}\left[1-\frac{3r}
{2(1+q)\sigmac}+ \frac{r^3}{2(1+q)^3 \sigmac^3}\right]} &
\sigmac<r<\sigmac +\sigmap \\
0 & r>\sigmac +\sigmap
\end{array}
\right.
\end{equation}
here $\zp$ is the fugacity of a pure ideal polymer system and
$q=\sigmap/\sigmac$.
In \fig{AO} we show the interaction potential for several size ratios $q$.
On decreasing $q$, the  attractive well potential becomes deeper 
and narrower.
Here we do not intend to describe in detail this model, which is well known 
in literature (see e.g.\ \cite{evansmol}), and was 
previously considered also by some of us \cite{HRT-F}. 
We just remind  that
this effective interaction disregards three-body and higher-body terms.
Interactions beyond the pair level are actually absent 
for $q=\frac{\sigmap}{\sigmac}<0.154$, but 
they are present for larger values of $q$, so that $v_{\rm AO}$ 
becomes less and less accurate
as $q$ increases.  
Despite of this limit and of its simplicity,
we stress 
that
the AO model
is able to reproduce also at a fairly quantitative level
the main features of the experimental phase diagrams
for a large range in the size ratios.
Finally, we recall that the fluid-fluid demixing becomes stable 
as $q$  increases and
the depletion potential becomes longer ranged \cite{gast,evansmol,lek}.

We applied HRT to the two-body AO potential \eq{ao} 
for different size ratios and we completed the theoretical analysis 
about the critical behavior of this kind of mixtures  preliminarily 
investigated in Ref.\ \cite{HRT-F}.
The theoretical results
are compared with  those
we obtained via Monte Carlo simulation and finite size scaling analysis
on the full mixture as well as on the effective pair interaction 
\eq{ao}.

\subsection{Hierarchical Reference Theory}

As introduced before, for the theoretical analysis we consider a model 
fluid of particles interacting via a pairwise additive potential $V({\bf 
r}_{1}, {\bf r}_{2}, \ldots {\bf r}_{n})= \sum_{i<j}v(|{\bf r}_{i}-{\bf 
r}_{j}|)$, where ${\bf r}_{i}$ is the position of a generic particle $i$. 
We assume that the two-body potential $v(r)$ is spherically symmetric and 
results from the sum of a very short-ranged, singular repulsion $v_{R}(r)$ 
and a longer-ranged attraction $w(r)$. In the present case $w(r)$ is 
$v_{\rm AO}(r)$, i.e. \eq{ao} for $r>\sigmac$.  The fluid interacting via 
$v_{R}(r)$ alone acts as the unperturbed or reference system, whose 
properties are considered as known. Here the reference system is the 
hard-sphere fluid, whose thermodynamics and correlations are accurately 
described by the Carnahan-Starling equation of state~\cite{hansen} and the 
Verlet-Weis parametrization of the two-body radial distribution 
function~\cite{verlet}. We then focus on the perturbation $w(r)$. In 
HRT~\cite{hrtrev}, $w(r)$ is switched on by taking its Fourier transform 
$\widetilde{w}(k)$ and introducing a parameter $Q$ and a potential 
$w_{Q}(r)$, such that its Fourier transform coincides with 
$\widetilde{w}(k)$ for $k>Q$, and is identically vanishing for $k<Q$. As 
$Q$ evolves from $Q=\infty$ to $Q=0$, the interaction 
$v_{Q}(r)=v_{R}(r)+w_{Q}(r)$ goes from the reference part $v_{R}(r)$ to 
the full potential $v(r)$. This procedure for turning on the interaction 
closely resembles that adopted in the momentum-space renormalization 
group~\cite{wilson,fisherlec}. At a given stage of the evolution, the 
nature of $w_{Q}(r)$ is such that fluctuations over length scales $L$ 
larger than $1/Q$ are suppressed, so that critical fluctuations are 
recovered only in the limit $Q\rightarrow 0$. The corresponding evolution 
of the Helmholtz free energy and $n$-body direct correlation functions, 
from the reference fluid to the fully interacting one, is described by an 
exact hierarchy of integral-differential equations. Close to a critical 
point and at large length scales, this hierarchy becomes indeed equivalent 
to a formulation of the momentum-space renormalization 
group~\cite{nicoll}. However, the HRT hierarchy remains valid also away 
from criticality and over all length scales, thereby describing also the 
nonuniversal behavior of the fluid, which depends on the specific features 
of the microscopic interaction.

The first equation of the hierarchy gives the evolution of the Helmholtz free
energy $A_{Q}$ of the partially interacting system in terms of its two-body 
direct correlation function in momentum space $c_{Q}(k)$ 
and the full perturbation $\widetilde{w}(k)$:
\begin{equation}
\frac{\partial {\cal A}_{Q}}{\partial Q}=-\frac{Q^2}{4\pi^2} \,
\ln \! \left(1-\frac{\Phi(Q)}{{\cal C}_{Q}(Q)}\right) \, .
\label{hrteq}
\end{equation}
In the equation above, we have set $\Phi(k)=-\widetilde{w}(k)/k_{\rm B}T$, 
$k_{\rm B}$ being the Boltzmann constant and $T$ the absolute temperature,
and the quantities ${\cal A}_{Q}$, ${\cal C}_{Q}(k)$ are linked to $A_{Q}$
and $c_{Q}(k)$ by the relations
\begin{eqnarray}
& & {\cal A}_{Q} = -\frac{A_{Q}}{k_{\rm B}T V}+\frac{1}{2} \rho^{2}
\left[\Phi(k\!=\!0)-\Phi_{Q}(k\!=\!0)\right]
-\frac{1}{2} \,
\rho\!\int\!\!\frac{d^{3}{\bf k}}{(2\pi)^{3}}
\left[\Phi(k)-\Phi_{Q}(k)\right]
\label{amod} \\
& & {\cal C}_{Q}(k) = c_{Q}(k)+\Phi(k)-\Phi_{Q}(k) \, ,
\label{cmod}
\end{eqnarray}
where $V$ is the volume and $\rho$ the number density. These modified free
energy and direct correlation function have been introduced in order to remove
the discontinuities which appear in $A_{Q}$ and $c_{Q}(k)$ at $Q=0$ and $k=Q$
respectively as a consequence of $\widetilde{w}_{Q}(k)$ itself being 
discontinuous at $k=Q$. Physically they represent the free energy 
and direct 
correlation function of the fully interacting fluid as given by a treatment
such that the Fourier components of the interaction with wavelengths larger 
than $1/Q$ are not  really disregarded, but 
instead they are approximately taken into account
by mean-field theory. 
In particular, for $Q=0$ the modified 
quantities coincide with the physical ones, once the fluctuations have been 
fully included,
while 
for $Q=\infty$ they give the mean-field expressions 
of the free energy and direct correlation function. 

In order to get a closed set of equations, 
\eq{hrteq} has been 
supplemented with a closure relation for ${\cal C}_{Q}(k)$. 
This is the point
where approximations are introduced into the HRT scheme. As in previous 
applications, our form of ${\cal C}_{Q}(k)$  has been inspired by  
liquid-state theories,  and it reads
\begin{equation}
{\cal C}_{Q}(k)=c_{\rm HS}(k)+\lambda_{Q}\,\Phi(k)+{\cal G}_{Q}(k)\, ,
\label{closure}
\end{equation}
where $c_{\rm HS}(k)$ is the direct correlation function of the hard-sphere
fluid, and $\lambda_{Q}$, ${\cal G}_{Q}(k)$ are {\it a priori} unknown 
functions of the thermodynamic state  and of $Q$. Specifically, 
the function 
${\cal G}_{Q}(k)$ is determined by the core condition, i.e., the requirement
that the radial distribution function $g_{Q}(r)$ be vanishing for every $Q$ 
whenever the interparticle separation is less than the hard-sphere diameter
$\sigma$.
$\lambda_{Q}$ is adjusted so that ${\cal C}_{Q}(k)$ satisfies
the compressibility rule. This constraint gives the reduced compressibility
of the fluid as the structure factor evaluated at zero wavevector, and can 
be expressed in terms of the modified quantities ${\cal A}_{Q}$, 
${\cal C}_{Q}(k)$ as
\begin{equation}
{\cal C}_{Q}(k\!=\!0)=\frac{\partial^{2}{\cal A}_{Q}}{\partial \rho^{2}} \, .
\label{sum}
\end{equation}
The compressibility rule~(\ref{sum}) plays a fundamental role in the present
scheme. In fact, when $\lambda_{Q}$ in \eq{closure} is determined via  
\eq{sum} and the resulting expression for ${\cal C}_{Q}(k)$ is used
in \eq{hrteq}, one obtains a partial differential equation for 
${\cal A}_{Q}$ which reads  
\begin{equation}
\frac{\partial {\cal A}_{Q}}{\partial Q}=-\frac{Q^2}{4\pi^2} \,
\ln \! \left[1-\frac{\Phi(Q)}{{\cal A}''_{Q} \varphi(Q) + \psi(Q)}\right] \, ,
\label{hrtclos}
\end{equation}
where we have set ${\cal A}''_{Q}=\partial^{2}{\cal A}_{Q}/\partial\rho^{2}$, 
$\varphi(k)=\Phi(k)/\Phi(0)$ and
\begin{equation}
\psi(k)=c_{\rm HS}(k)+{\cal G}_{Q}(k)-[c_{\rm HS}(0)+{\cal G}_{Q}(0)]
\varphi(k) \, .
\label{psi}
\end{equation}
\eq{hrtclos} is integrated numerically from $Q=\infty$ down to $Q=0$. 
At each integration step, ${\cal G}_{Q}(k)$ is determined by 
the core condition $g_{Q}(r)=0$, $0<r<\sigma$. This condition acts as 
an auxiliary equation for determining $\psi(k)$ via \eq{psi}. 
The function ${\cal G}_{Q}(k)$ has been approximated by a fourth-degree
polynomial in the interval $0<r<\sigma$, and the equations for the coefficients
were obtained. In order to keep the computational scheme relatively simple, 
further approximations were introduced in these auxiliary equations, 
which amount to decoupling the short- and long-range part of the correlations.
The details of this procedure have been given elsewhere~\cite{mero}.

Since in this work we will be mostly concerned with the critical region, it is
worthwhile recalling the critical behavior of HRT with the closure 
relation~(\ref{closure})~\cite{hrtrev,hrt1}. In the critical region 
and at small $Q$, 
\eq{hrtclos} can be considerably simplified in such a way that 
it depends just on the long-wavelength limit of the direct correlation 
function. This in turn is determined by the compressibility rule~(\ref{sum})
and the assumption, implicit in \eq{closure}, that ${\cal C}_{Q}(k)$ 
is always an analytic function of $k$:
\begin{equation}
{\cal C}_{Q}(k)\sim \frac{\partial^{2}{\cal A}_{Q}}{\partial \rho^{2}} 
- b k^{2} \, ,
\label{smallk}
\end{equation} 
where $b$ is a regular function of $Q$, $\rho$, $T$. The resulting evolution
equation is then cast into universal form by suitably rescaling the density 
and the free energy.
The critical behavior of the theory is analyzed
in terms of fixed-point functions and linearized flow of the rescaled 
free energy in the neighborhood of the fixed points, along the lines 
of the renormalization group~\cite{hrt1}. The resulting critical exponents
are correct to first order in the expansion in the parameter $\epsilon=4-d$,
$d$ being the dimensionality of the system. In particular, the analytic 
dependence of 
the direct correlation function on $k$, also known as the Ornstein-Zernike 
{\it ansatz}, implies that the critical exponent $\eta$ is zero 
in our approximation. For $d=3$ one finds~\cite{hrt1} $\gamma=1.378$, 
$\beta=0.345$, $\delta=5$, $\alpha=-0.07$, $\nu=\gamma/2=0.689$, 
where the usual notation
for the critical exponents has been used. These exponents satisfy 
the algebraic relations implied by the scaling of the free energy 
in the critical region~\cite{fisherlec}. Below the critical temperature, 
the theory correctly predicts a diverging compressibility inside 
the coexistence region. However, the compressibility diverges also 
on the coexistence boundary, unlike in the real fluid. This is a consequence
of the Ornstein-Zernike ansatz~(\ref{smallk}). 

Finally, we observe that, for the AO fluid considered here, 
the perturbation $\Phi(k)$ does not depend on temperature. The role 
of the inverse temperature is instead played by the packing fraction 
of the polymer in the reservoir $\etapr$, defined in terms of the polymer
fugacity $\zp$ as $\etapr = \pi \zp \sigmap^3 / 6$ (see \eq{ao}). 
For $\etapr$ above a critical value $\etaprcr$, the AO model will phase separate
into a colloid poor phase (the colloidal vapor, with colloid packing fraction
$\etacg$) and colloid rich phase (the colloidal liquid, with colloid packing
fraction $\etacl$). In this sense, then, $\etapr$ is the analogue of inverse
temperature in fluid-vapor transitions of simple fluids. 
Both the inverse temperature and $\etapr$ appear in \eq{hrtclos} 
just as external parameters which govern the strength of the interaction
for an atomic fluid and the AO fluid respectively.
Consequently, this does not imply 
any substantial change in our treatment (see \olcite{HRT-F}).

\section{Simulations}
\label{SIM}

To test the predictions of HRT, large-scale Monte Carlo (MC) simulations 
of the AO model have also been performed, using colloid-to-polymer size 
ratios $q=0.8$, $q=0.56$ and $q=0.4$. The simulations were carried out in 
the grand canonical ensemble. In this ensemble, the volume $V$, the 
colloid fugacity $\zc$, and the polymer fugacity $\zp$ are fixed, while 
the number of particles fluctuates. We use cubic simulation boxes with 
edge $L$ and periodic boundary conditions in all $d=3$ dimensions. The 
output of the simulations consists of the distribution $\pc$, defined as 
the probability of observing a system containing $\nc$ colloids at 
``temperature'' $\etapr$, colloid fugacity $\zc$ and box size $L$.

The simulations of the {\it full} AO model, in which {\it both} the 
colloids {\it and} the polymers are explicitly retained, were performed 
using the method of \olcite{vink.horbach:2004*1}. The essential 
ingredients of this approach are a cluster move \cite{vink.horbach:2004*1, 
vink.horbach:2004}, a reweighting scheme \cite{virnau.muller:2004}, and 
histogram extrapolation \cite{ferrenberg.swendsen:1988, 
ferrenberg.swendsen:1989}. In addition, we performed a number of grand 
canonical simulations using the effective pair potential of \eq{ao}. These 
simulations, obviously, do not require the cluster move of 
\olcite{vink.horbach:2004*1}, and were performed using standard grand 
canonical MC \cite{frenkel.smit:2001, landau.binder:2000}.

One aim of the simulations is to verify to what degree the critical 
properties of the AO model predicted by HRT are reproduced. Previous 
simulations have shown that the AO model belongs to the 3D Ising 
universality class \cite{vink.horbach:2004*1, vink.horbach:2004, 
vink.horbach.ea:2004}, and that pronounced deviations from mean-field 
behavior become visible upon approach of the critical point. One strong 
point of HRT lies in its ability to yield non-classical (i.e.~non 
mean-field) critical exponents. The latter is expected to resemble more 
closely simulations, and indeed experiments \cite{chen.payandeh:2000, 
chen.payandeh:2001, royall}. In this work, we are particularly concerned 
with the critical behavior of the order parameter and the compressibility 
in the one-phase region. More precisely, defining $t \equiv 
\etapr/\etaprcr-1$ as distance from the critical point, the order 
parameter $\Delta \equiv (\etacl - \etacg)/2$ is expected to obey $\Delta 
= A t^\beta$, while the (dimensionless) compressibility in the one-phase 
region $\chi \equiv \vc (\avg{\nc^2} - \avg{\nc}^2) / V$ is expected to 
diverge as $\chi = B (-t)^{-\gamma}$, with critical exponents $\beta$ and 
$\gamma$, critical amplitudes $A$ and $B$, and $\vc = \pi \sigmac^3 / 6$ 
the volume of a single colloid. Note that this definition of $\chi$ 
differs from the one adopted in \olcite{vink.horbach.ea:2004} by a factor 
$1/\vc$. The quantity $\chi$ is related to the usual reduced 
compressibility $\chired$, i.e., the zero-$k$ value of the structure 
factor, by $\chi=\etac \, \chired$. Preferred values of the critical 
exponents for the 3D Ising universality class are listed in 
\tab{3dexp}.

\begin{table}[b]
\caption{\label{3dexp} Preferred values of the critical exponents of the 
specific heat $(\alpha)$, order parameter $(\beta)$, compressibility 
$(\gamma)$, and correlation length $(\nu)$ for the 3D Ising model 
\cite{fisher.zinn:1998}.}
\begin{ruledtabular}
\begin{tabular}{cccc}
$\alpha$ & $\beta$ & $\gamma$ & $\nu$ \\ \hline
0.109 & 0.326 & 1.239 & 0.630 \\
\end{tabular}
\end{ruledtabular}
\end{table}

To accurately probe the critical properties, the simulation data is analyzed
using a number of finite size scaling (FSS) techniques, which we will briefly
outline in what follows. Most notably, part of our analysis is based on recently
proposed {\it unbiased} scaling algorithms \cite{kim.fisher.ea:2003,
kim.fisher:2004}. We emphasize here that all FSS algorithms require as input
highly accurate MC data, typically for a range of system sizes and temperatures.
Since our resources are limited, some compromise is unavoidable. The FSS
analysis in this work is therefore limited to the full AO model only; no FSS is
performed on the data obtained using the effective pair potential of \eq{ao}.

\subsection{Cumulant Intersections (CI)}
\label{CI}

The cumulant intersection approach \cite{binder:1981} is a common FSS 
method to determine the critical temperature $\etaprcr$ from simulation 
data. Here, the grand canonical distribution $\pc$ is used to measure the 
cumulant ratio $\avg{m^2} / \avg{|m|}^2$ with $m = \nc - \avg{\nc}$ as 
function of $\etapr$, for a number of system sizes $L$. In these 
simulations, the colloid fugacity is tuned so as to obey the 
``equal-area'' criterion \cite{binder.landau:1984, borgs.kotecky:1992}. 
The data from the different system sizes is expected to show a common 
intersection point, which yields an estimate of the critical polymer 
reservoir packing fraction $\etaprcr$. For the full AO model with $q=0.8$, 
the results of this procedure can be found in 
\olcite{vink.horbach:2004*1}. Additional simulations for $q=0.56$ 
performed in this work, using three distinct system sizes $L=10,12,14$ (in 
units of the colloid diameter), show qualitatively similar behavior. The 
resulting estimates of $\etaprcr$ are listed in \tab{summary}. Note that 
the cumulant intersection approach can be applied without prior knowledge 
of the universality class. In this sense, then, it is an unbiased method.

The cumulant intersection method thus requires as input MC data from at 
least two different system sizes. To verify the consistency of the 
results, however, a higher number is recommended. For $q=0.4$, this 
requirement exceeded the computational resources available to us, and so 
no cumulant intersection result for this size ratio is reported (small 
size ratios $q$ are problematic to simulate due to the high number of 
polymers these simulations require). The results for $q=0.4$ are 
determined by means of the economical scaling algorithm and scaling plots 
to be described shortly.

\subsection{Scaling Plots (SP)}

According to finite size scaling theory, the order parameter $\Delta$ 
obtained in a finite system of linear dimension $L$ close to criticality 
shows a systematic $L$ dependence that can be written as $\Delta = 
L^{-\beta/\nu} {\cal M}^0 (t L^{1/\nu})$, with $\nu$ the critical exponent 
of the correlation length, and ${\cal M}^0$ a scaling function independent 
of system size \cite{landau.binder:2000, newman.barkema:1999}. The latter 
implies that plots of $L^{\beta/\nu} \Delta$ versus $t L^{1/\nu}$ should 
collapse onto a single curve. For large $t L^{1/\nu}$, but still within 
the critical region, this curve should approach the critical power law of 
the thermodynamic limit. The latter is most conveniently visualized using 
a double logarithmic scale. The data is then expected to approach a 
straight line, with slope $\beta$ and intercept equal to the critical 
amplitude $A$. This method can thus be used to extract critical amplitudes 
from simulation data. Note that this approach is biased, in the sense that 
the critical temperature and exponents must be known {\it a priori}. A 
possible strategy is to obtain the critical temperature using the cumulant 
intersection method, and to simply assume 3D Ising universality. For 
many fluids, especially those with short-ranged interactions, the latter 
assumption will be a safe one. Precisely this strategy was followed in 
\olcite{vink.horbach.ea:2004} to obtain the critical amplitude of the 
order parameter, as well as the compressibility amplitude, for $q=0.8$. 
The scaling plots obtained by applying the same strategy to the $q=0.56$ 
data of this work are qualitatively similar; the resulting estimates of 
the critical amplitudes are listed in \tab{summary}. Note that in these 
analysis, the colloid fugacity was again chosen to fulfill the 
``equal-area'' criterion (see section \ref{CI}), and also that, for the 
same reason as before, no estimates for $q=0.4$ are reported.

\subsection{Unbiased Scaling (US)}

Recently, FSS algorithms were presented that do not require prior 
knowledge of the critical exponents, nor of the critical temperature 
\cite{orkoulas.fisher.ea:2001, kim.fisher:2003, 
kim.fisher.ea:2003,kim.fisher:2004, kim:2005}. Instead, these quantities 
are {\it outputs}, and this may prove valuable if there is doubt 
regarding the universality class of a system. In fact, these algorithms 
were inspired by serious doubts raised over the universality class of the 
restricted primitive electrolyte (which was shown to be that of the 3D 
Ising model \cite{luijten.fisher.ea:2002}). In case of the AO model, there 
is no doubt regarding the universality class. The motivation for 
nevertheless using these new unbiased FSS methods is a different one. As 
mentioned in \olcite{kim.fisher.ea:2003}, one problem in simulating 
asymmetric fluids lies in choosing the coexistence chemical potential (or 
fugacity). A common approach, also adopted by us before, is to use the 
``equal-area'' criterion: the fugacity (of the colloids) is chosen such 
that $\pc$ becomes bimodal, with two peaks of equal area. Away from the 
critical point, the peaks in $\pc$ are well separated: equal area then 
corresponds to equal pressure in the two phases 
\cite{orkoulas.fisher.ea:2001}, a necessary condition for phase 
coexistence. Close to the critical point, however, the peaks in $\pc$ 
strongly overlap, and a separation in terms of equal area becomes rather 
arbitrary. An alternative procedure is thus desirable.

\begin{figure}
\begin{center}
\includegraphics[width=\figwidth,clip]{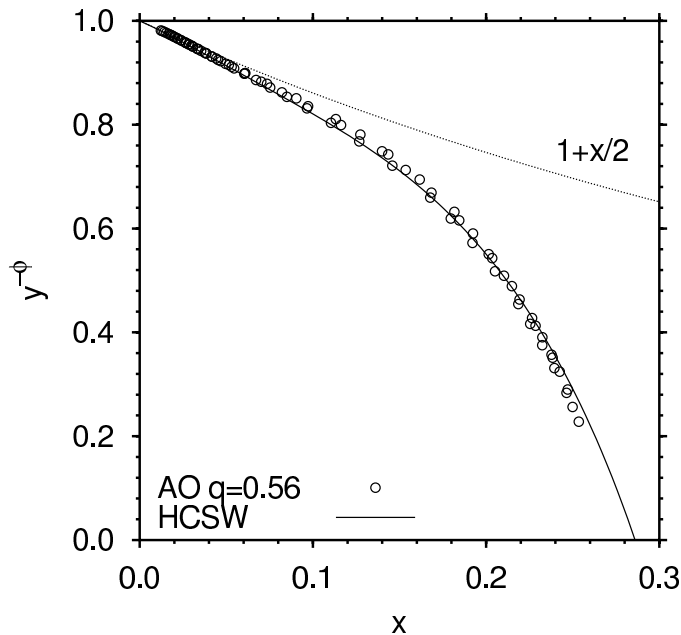}
\caption{\label{scaling} Scaling function of the order parameter obtained 
using the unbiased scaling algorithm. Following the convention of 
\olcite{kim.fisher:2004}, the scaling function is raised to a negative 
exponent, with $\phi=1/\beta$. Open circles show results obtained in this 
work for the AO model with $q=0.56$; the solid curve is the HCSW result of 
\olcite{kim.fisher:2004}. Also shown is the exact small $x$ limiting form 
$y = 1 + x/2$.}
\end{center}
\end{figure}

The unbiased algorithm \cite{kim.fisher.ea:2003, kim.fisher:2004} requires 
as input the grand canonical distribution $\pc$ for at least three 
different system sizes $L$, and for $\etapr$ ranging from the non-critical 
regime toward the critical point. It is therefore considerably more 
expensive than previously discussed methods, which only required data near 
the critical point. For this reason, we consider $q=0.56$ only. Starting 
with $\etapr$ far away from the critical point in the two-phase region, 
the cumulant ratio $\avg{m^2}^2 / \avg{m^4}$ is plotted as function of the 
average colloid packing fraction $\vc \avg{\nc}/V$, with symbols defined 
as before (note that this plot is parameterized by $\zc$). The resulting 
curve will reveal two minima, located at $\etac^-$ and $\etac^+$, with 
respective values $Q^-$ and $Q^+$ at the minima. Defining the quantities 
$Q_{\rm min} = (Q^+ + Q^-)/2$, $x = Q_{\rm min} \ln (4/e Q_{\rm min} )$, 
and $y = (\etac^+ - \etac^-) / (2 \Delta)$, the points $(x,y)$ from the 
different system sizes should all collapse onto the line $y=1+x/2$. Recall 
that $\Delta$ is the order parameter in the thermodynamic limit at the 
considered $\etapr$, precisely the quantity of interest, which may thus be 
obtained by fitting until the best collapse onto $1+x/2$ occurs. In the 
next step, $\etapr$ is chosen closer to the critical point, the points 
$(x,y)$ are calculated as before, but this time $\Delta$ is chosen such 
that the new data set joins smoothly with the previous one, yielding an 
estimate of the order parameter at the new temperature. This procedure is 
repeated all the way to the critical point, where $\Delta$ vanishes, 
leading to an estimate of $\etaprcr$. Moreover, the procedure also yields 
$y$ as function of $x$. The latter function is universal within a 
universality class, and for the hard-core square-well (HCSW) fluid can be 
found in \olcite{kim.fisher:2004}. Since the AO model belongs to the same 
universality class, we should arrive at a similar result. The latter is 
verified in \fig{scaling}, which shows $y$ as function of $x$ obtained in 
this work, compared to the result of \olcite{kim.fisher:2004}. Note that, 
for small $x$, our data correctly approach $y=1+x/2$. More importantly, 
the scaling function we obtain agrees well with the one obtained in 
\olcite{kim.fisher:2004}, albeit that our data is ``noisier''. This comes 
as no surprise, since the HCSW fluid is much easier to simulate than the 
AO model, and so higher quality data are more readily generated. The 
critical value $\etaprcr$ is obtained via the best collapse of 
$(t,\Delta)$ onto a power law. The resulting data is plotted in \fig{FIG3} 
(triangles), and the corresponding estimate of $\etaprcr$ is listed in 
\tab{summary}. The critical exponent $\beta$ and amplitude $A$ are 
obtained from the slope and intercept of the simulation data. For the 
exponent we find $\beta \approx 0.32$, which is in good agreement with the 
accepted 3D Ising value; the critical amplitude is listed in 
\tab{summary}.

\begin{figure}
\begin{center}
\includegraphics[width=\figwidth,clip]{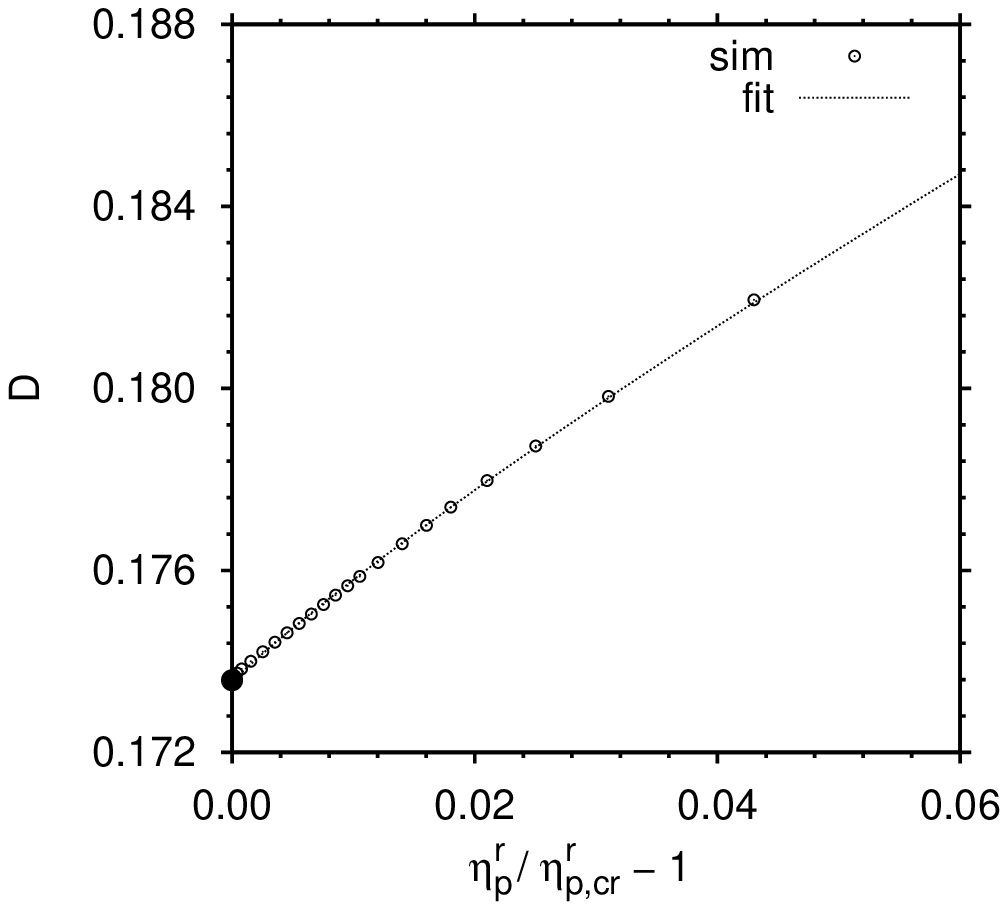}

\caption{\label{coex} Coexistence diameter of the $q=0.56$ AO model as 
function of the relative distance from the critical point, where 
$\etaprcr=0.6256$ was used. Open circles show simulation results, 
the closed circle is our estimate of $\etaccr$ obtained by fitting the 
simulation data to \eq{eq:coex}; the curve shows the fit itself.}

\end{center}
\end{figure}

The unbiased scaling algorithm naturally extends to estimate the 
coexistence diameter $D \equiv (\etacl+\etacg)/2$ from simulation data 
\cite{kim.fisher.ea:2003, kim:2005}. In contrast to the order parameter, 
however, the corresponding scaling function for the diameter is {\it not} 
universal. In other words, a comparison to \olcite{kim:2005} similar in 
spirit to \fig{scaling}, is not possible in this case. Instead, we simply 
show in \fig{coex} the output of the scaling algorithm for the $q=0.56$ AO 
model, where we used for $\etaprcr$ the value obtained in the previous 
paragraph. Close to the critical point, the diameter is expected to scale 
as
\begin{equation}
\label{eq:coex}
 D = \etaccr \left( 1 + a_1 t^{2\beta} + a_2 t^{1-\alpha} + a_3 t \right), 
\end{equation}
with $t$ the relative distance from the critical point, $\etaccr$ the 
critical colloid packing fraction, and non-universal amplitudes $a_i$ 
\cite{kim.fisher.ea:2003*b}. The curve in \fig{coex} is a fit to the data 
using the above form yielding $\etaccr=0.1736$, $a_1= -0.125$, $a_2 = 
1.674$ and $a_3=-0.875$. The HRT diameter (not shown in the figure) varies 
more slowly with $t$, its average slope being about half of the 
simulation results in the reduced temperature interval shown in 
\fig{coex}.

\subsection{Economical Scaling (ES)}

The authors of the unbiased scaling algorithm have also presented, in 
\olcite{kim.fisher:2004}, a biased version of their algorithm. If one is 
prepared to accept 3D Ising universality, the scaling function of the 
HCSW fluid shown in \fig{scaling}, which is universal within a 
universality class, can be used to estimate the order parameter of other 
systems in that class. To use this approach, MC data of a single system 
size obtained close to criticality is in principle sufficient. Since we 
lack the resources to execute the full unbiased scaling algorithm for 
$q=0.4$ and $q=0.8$, this economical approach offers an attractive 
alternative. For $q=0.4$, a single simulation was thus performed using 
system size $L/\sigmac=10$, while for $q=0.8$ the data of the largest 
system of \olcite{vink.horbach:2004*1} was used. The triangles in 
\fig{FIG4} and \fig{FIG2} show the corresponding order parameter as 
function of the distance from the critical point. Here, $\etaprcr$ was 
obtained from the best collapse of $(t,\Delta)$ onto a power law; this 
also yields the critical amplitude $A$. Using the value of $\etaprcr$ thus 
obtained for $q=0.4$, a scaling plot was generated to also extract the 
critical amplitude $B$. The latter estimate will not be very precise, but 
seems consistent at least with the trend that smaller size ratios lead to 
smaller compressibility amplitudes.

Note that at present, no economical algorithm for the coexistence diameter 
exists, since the corresponding scaling function is {\it not} universal in 
this case \cite{kim:2005}. Hence, in \tab{summary}, no FSS estimate of 
$\etaccr$ could be provided for $q=0.4$. Instead, we have listed the 
colloid packing fraction at $\etaprcr$ in the finite system. For $q=0.8$, 
the estimate of \olcite{vink.horbach:2004} is reported, which was derived 
using the Bruce-Wilding mixed-field scaling method 
\cite{bruce.wilding:1992}.

\begin{table*}[b]

\caption{\label{summary} Summary of our FSS results for the full mixture 
AO model at various size ratios $q$. Listed are the critical {\it 
temperature} $\etaprcr$, the critical amplitude $A$ of the order 
parameter, the critical amplitude $B$ of the compressibility in the 
one-phase region, and the critical colloid packing fraction $\etaccr$. 
Also indicated is the FSS method that was used to obtain the estimate: 
cumulant intersection (CI), scaling plot (SP), unbiased scaling (US), and 
economical scaling (ES). The $\star$ symbol marks the estimate we believe 
to be the most reliable, in case multiple values are provided.}

\begin{ruledtabular}
\begin{tabular}{ccccc}
$q$ & $\etaprcr$ & $A$ & $B$ & $\etaccr$ \\ \hline
0.8  & $0.766 \pm 0.002$ (CI) & $0.27 \pm 0.02$ (SP) & $0.053 \pm 0.002$
(SP) & $0.1340 \pm 0.0006$\footnote{Taken from \olcite{vink.horbach:2004}
and listed here for completeness.} \\
     & $0.7646 \pm 0.0003^\star$ (ES) & $0.28 \pm 0.01^\star$ (ES) & & \\ \hline
0.56 & $0.6258 \pm 0.0002$ (CI) & $0.38 \pm 0.01$ (US) & $0.049 \pm 0.002$ (SP) & $0.1736 \pm 0.0004$ (US) \\
     & $0.6256 \pm 0.0001^\star$ (US) & & & \\ \hline
0.4  & $0.52154 \pm 0.0001$ (ES) & $0.47 \pm 0.01$ (ES) & $0.045 \pm 0.003$ (ES+SP) & $\sim 0.21$\footnote{This is no FSS estimate!}
\end{tabular}
\end{ruledtabular}

\end{table*}

\begin{table}[b]

\caption{\label{hrt} Critical {\it temperature} $\etaprcr$ and critical 
colloid packing fraction $\etaccr$ determined with HRT and mean-field (MF) 
theory for the different size ratios.}

\begin{ruledtabular}
\begin{tabular}{cccc}
 & q& $\etaprcr$ & $\etaccr$  \\ \hline
HRT & 0.8& 0.4825 & 0.1895  \\ \hline 
 &0.56& 0.4679 & 0.2204   \\ \hline
& 0.4 &  0.4404 & 0.2257    \\ \hline \hline 
MF & 0.8&    0.4129  &        0.1304  \\ \hline
 &  0.56&  0.4086   &    0.1304            \\\hline
 & 0.4&  0.3718    &  0.1304               \\
\end{tabular}
\end{ruledtabular}

\end{table}

\section{Results and discussion}

As introduced before, in \fig{AO} we present the pair interaction for 
the different size ratios we focused on, i.e. $q=0.4,0.56,0.8$.
\begin{figure}
\begin{center}
\includegraphics[width=\figwidth,clip]{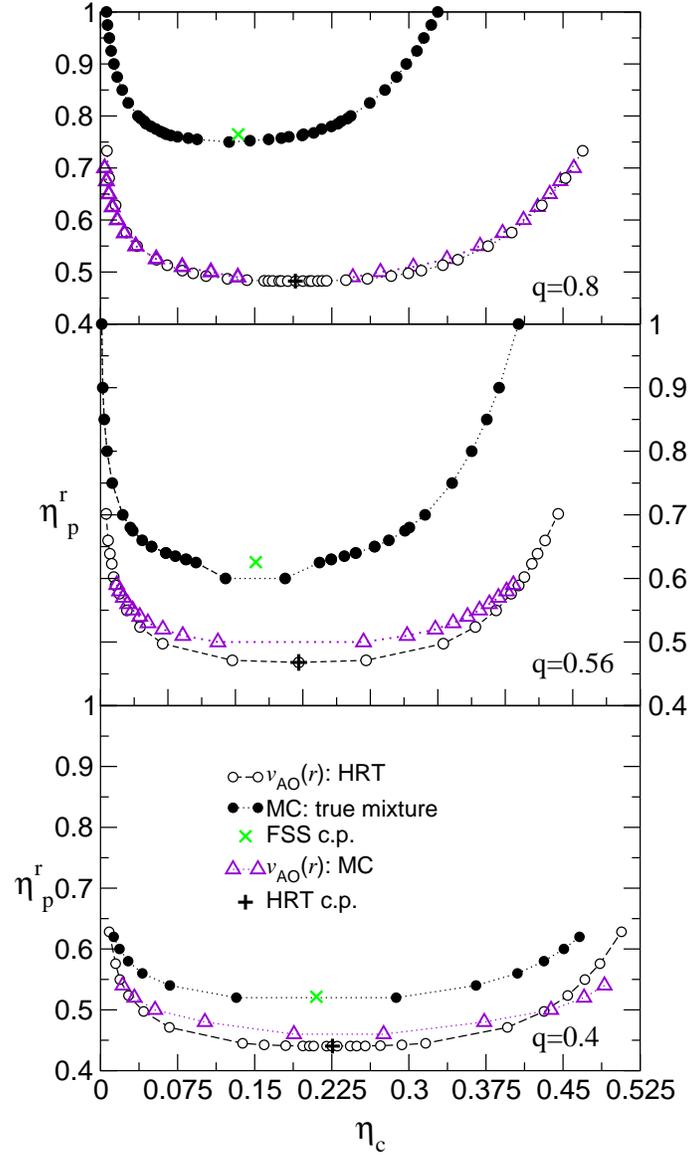}
\caption{\label{FIG1} Coexistence curve of the AO model for several colloid-to-polymer size ratios
$q$. Closed circles are simulation results of the full mixture description, where the
crosses mark the location of the critical point obtained using FSS. The triangles are
simulation results obtained using the effective pair potential of \eq{ao}. The open circles
are the theoretical predictions obtained using HRT. Lines connecting the points serve to
guide the eye.
The agreement between the pair interaction and the full-mixture picture
increases on decreasing the size ratio.
On the contrary, the accuracy of the HRT with respect to the simulation 
results, both on $v(r)$, increases on increasing $q$.}
\end{center}
\end{figure}
Starting from the effective interaction which describes the AO mixtures,
we studied the critical behavior by HRT
and compared our results with simulations on the same effective interaction.
In addition, we compared our analysis of the critical behavior with 
simulations performed on the full, two-component mixture.

We first considered the whole coexistence region in order to assess the limits
entailed by a description of this mixture in terms of an effective pair 
interaction as well as by the particular HRT scheme which we used. 
In Ref. \cite{HRT-F}, some of us showed that HRT yields a very good account 
of fluid-fluid coexistence curve 
and of its stability with respect to solid phases for several $q$ ($q=0.25,0.4,0.6,0.8$).
Thanks to its renormalization-group structure, the accuracy of HRT 
in the critical region is remarkable compared to
other approximate theories, such as integral equations 
or perturbative methods.
In \fig{FIG1} we show both the data we obtained with HRT and MC simulation 
on the pair potential $v_{\rm AO}(r)$, and the results of our study 
with MC simulation on the full mixture.
There are two main features which emerge from the comparison: on the one hand,
the agreement between theory and simulation results on the pair interaction
is quite good, and increases on increasing the size ratio, because of the
theory being tailored, at the present stage, to relatively long-ranged
interactions 
(the apparent discrepancy in this trend, central panel, is due
to the smaller number of MC data close to $\etaccr$
for $q=0.56$ than for $q=0.4$). On the other hand, the agreement between 
the effective pair interaction description and the full binary mixture 
increases 
on decreasing the size ratio, both for the overall coexistence curve and
the critical packing fractions $\etaprcr$ and $\etaccr$. 
This is due to a deficiency of the effective pair potential,
which is more appropriate to describe colloid-polymer mixtures for small $q$.
In general, the full AO mixture appears to be much more sensitive to a change
in $q$ than its one-component representation in terms of the effective
pair interaction $v_{\rm AO}(r)$. This is especially true for $\etaprcr$.
In fact, most of the deviations of the HRT critical parameters 
from the simulation results of the full mixture are not due to the HRT 
approximation, but to the modeling of the mixture in terms of a pair 
interaction, as one can appreciate by comparing the HRT coexistence curves 
with the simulation results for the pair interaction model. In \tab{hrt},
the HRT results for the critical packing fractions $\etaprcr$ and $\etaccr$ 
are compared to the mean-field values. 
We recall that the mean-field approximation is obtained by setting
$Q=\infty$ in \eq{amod}. This corresponds to evaluating the excess free
energy with respect to the hard-sphere gas {\em a la} van der Waals as
$A-A_{\rm HS}=\frac{1}{2}\rho^{2}V\int \!\! d^{3}{\bf r}\, v_{\rm AO}(r)$.

\begin{figure}     
\begin{center}  
\includegraphics[width=\figwidth,clip]{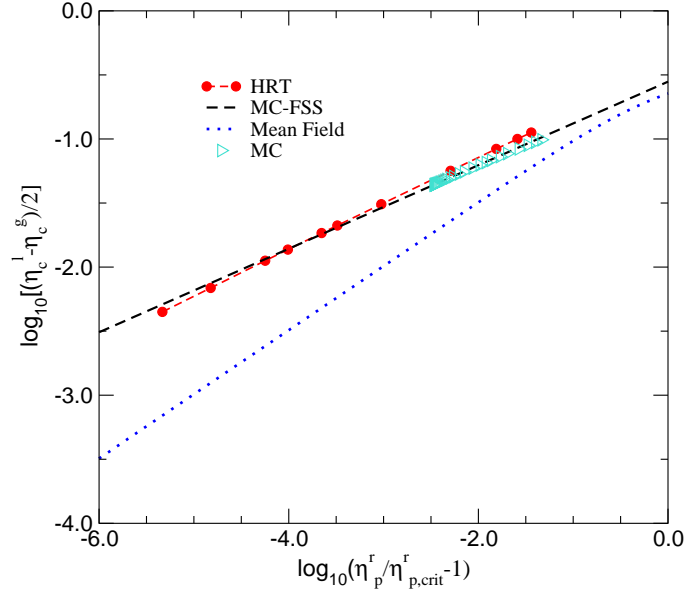}
\caption{\label{FIG2} Coexistence curve for $q=0.8$ close to the critical point on double logarithmic scales
(reduced units). The quantities $\etacg$ and $\etacl$ are the colloid packing fraction on the low- and high density branch of the coexistence curve respectively.
The triangles are data obtained in MC
simulations of the full mixture AO model extrapolated to the thermodynamic limit using FSS.
The dashed line is a fit to the simulation data using the critical power law $\Delta = A
(\etapr/\etaprcr-1)^\beta$, assuming the 3D Ising value for $\beta$, and fit parameters $\etaprcr$ and $A$
taken from \tab{summary}. Closed circles show the HRT result; the dotted line represents the
mean-field result.}
\end{center}
\end{figure}

The main purpose of this paper is to study in detail 
the critical behavior of the AO system
with simulations and theory 
and to test the accuracy
of HRT in reproducing  nontrivial critical exponents.
The discrepancy between the HRT critical point and the value obtained via
finite-size scaling on the AO binary mixture is then at least partially
accounted for by adopting the reduced ``temperature'' variable
$|\etapr/\etaprcr-1|$, as customary in the study of the critical behavior.
Moreover, such a discrepancy is not expected to affect the universal features
of the transition.
\begin{figure}
\begin{center}
\includegraphics[width=\figwidth,clip]{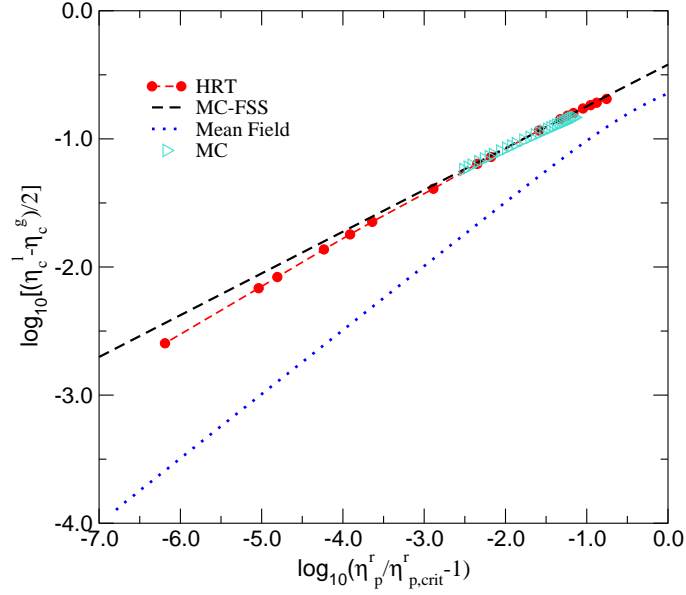}
\caption{\label{FIG3} Coexistence curve for $q=0.56$ (reduced units). 
Symbols
defined as in \fig{FIG2}.}
\end{center}
\end{figure}
\begin{figure}
\begin{center}
\includegraphics[width=\figwidth,clip]{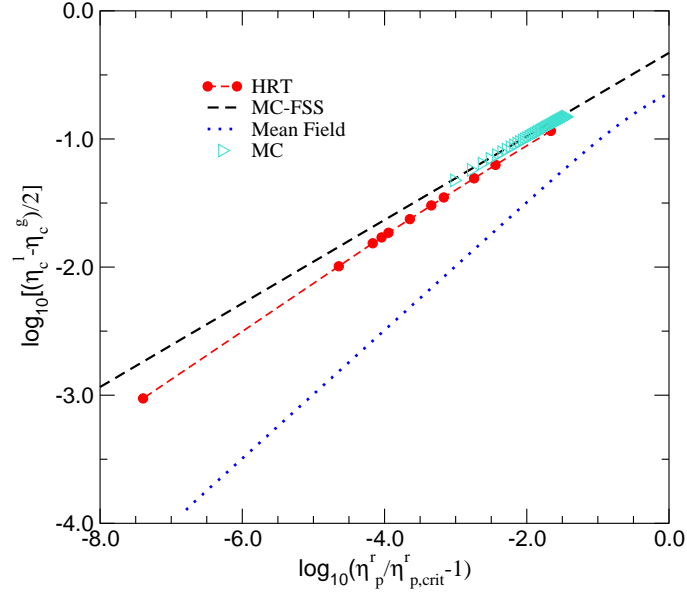}
\caption{\label{FIG4} Coexistence curve for $q=0.4$ close to the critical point
(reduced units).
Symbols
defined as in \fig{FIG2}.}
\end{center}
\end{figure}
\begin{figure}
\begin{center}
\includegraphics[width=\figwidth,clip]{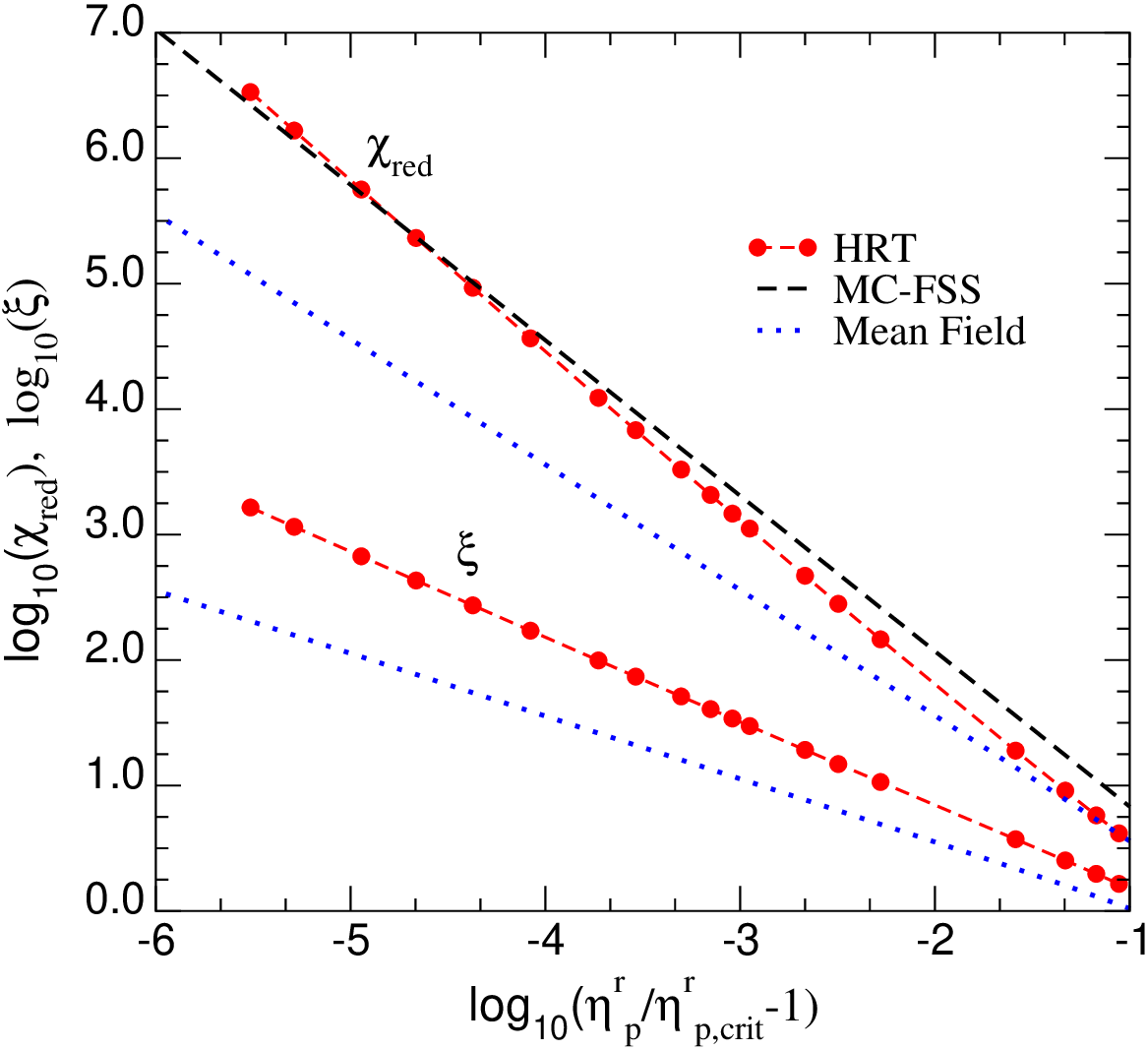}%
\caption{\label{FIG5} Reduced compressibility $\chired$ (upper curves) and correlation length $\xi$ (lower curves)
for the AO fluid with $q=0.8$.}
\end{center}
\end{figure}
We then focus on the critical behavior of the AO model,
in particular on the study of the order parameter, compressibility and correlation function.
In \fig{FIG2} we present 
the HRT data for the difference between the reduced
densities of the colloid on the ``liquid'' and ``vapor'' branches
of the coexistence curve
$\Delta=(\eta_{c}^{l}-\eta_{c}^{v})/2$
as a function of the reduced ``temperature''
$\etapr/\etaprcr-1$
for $q=0.8$. A linear fit of our results
for $\log(t)\lesssim -4$ gives a power-law behavior
with $\beta=0.37$.
We recall that the asymptotic
value of the exponent $\beta$ predicted by HRT  is obtained
by linearizing the renormalization-group flow induced by the HRT evolution
equation (\ref{hrteq}) in the neighborhood of the critical point and 
it is given by $\beta=0.345$ \cite{hrtrev}.
The theoretical results are compared
with the asymptotic Ising-3D
power law $\Delta\sim A\, t^{\beta}$ obtained from finite-size
scaling. The critical exponent $\beta$ of the finite-size scaling analysis
coincides with the generally accepted value $\beta=0.324$ 
(see Sec.\ \ref{SIM}). 
The same figure shows the data obtained by simulating  the full mixture.
The comparison with the mean-field results (dotted curve)
evidences how HRT is able to reproduce
the non-trivial critical behavior of the mixture.
In \fig{FIG3} and \fig{FIG4} we present the same analysis
for different size ratios.
According to both HRT and finite-size scaling, the critical amplitude
of the coexistence region increases on decreasing $q$ as a consequence
of the coexistence curve becoming flatter (see \fig{FIG1}). However, this
trend is stronger for the binary system than for the one-component fluid,
as can be inferred by comparing the values reported in \tab{summary} 
with the amplitudes obtained by the linear fit of the HRT results   
$A=0.41$, $A=0.52$, $A=0.56$ for $q=0.8$, $q=0.56$, 
and $q=0.4$ respectively.  
As a consequence, in the right-hand side of the reduced temperature axis,
HRT overestimates the finite-size scaling data for $q=0.8$, while
it underestimates them for $q=0.4$. The overall agreement remains
nevertheless quite good, especially if one considers that the theory contains
no free parameters.
Far from the critical point we recover the mean-field trend.

\begin{figure}
\begin{center}
\includegraphics[width=\figwidth,clip]{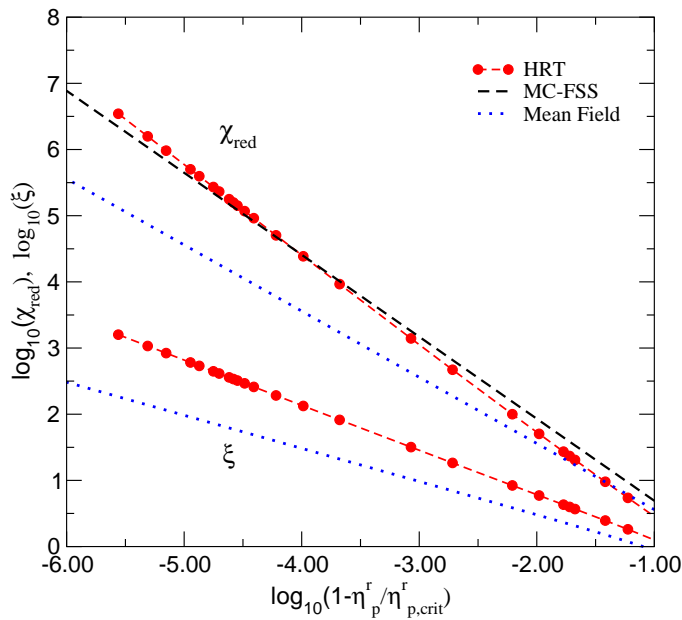}
\caption{\label{FIG6} Reduced compressibility $\chired$ (upper curves) and correlation length $\xi$ (lower curves)
for the AO fluid with $q=0.56$.}
\end{center}
\end{figure}

\begin{figure}
\begin{center}
\includegraphics[width=\figwidth,clip]{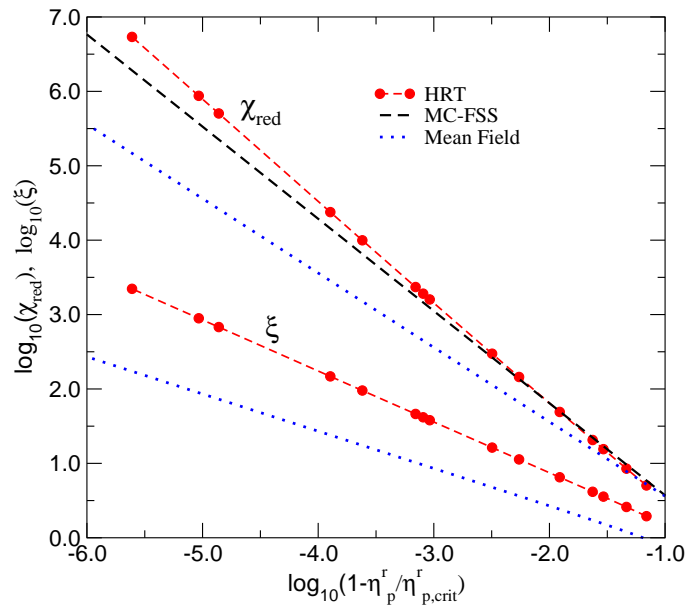}
\caption{\label{FIG7} Reduced compressibility $\chired$ (upper curves) and correlation length $\xi$ (lower curves)
for the AO fluid with $q=0.4$.}
\end{center}
\end{figure}

In \fig{FIG5}, \fig{FIG6} and \fig{FIG7}
 we consider the behavior of the reduced compressibility
$\chi_{\rm red}$ of the colloid on the critical isochore in the
one-phase region. We report both the HRT results and the asymptotic
power law obtained from finite-size scaling,
$\chi_{\rm red}\sim B^{\prime}|t|^{-\gamma}$, 
$\gamma=1.239$ \cite{vink.horbach.ea:2004},
$B^{\prime}=B/\etaccr$ with $B$
given in \tab{summary}.
We recall the HRT result:
$\gamma=1.378$, about $10\%$ larger
than the correct one \cite{hrtrev}.
A linear fit of the data shown in the figures 
for $\log(t)\lesssim -4$ gives $\gamma=1.37$.
The effect of the interaction range on the critical amplitude
of the compressibility is weaker than on the order parameter, and is again
more evident in the finite-size scaling results for the binary mixture
than in the one-component fluid. In fact, the data reported in \tab{summary}
show a weak decrease of the critical amplitude on decreasing $q$, while
the amplitude of the HRT compressibility is hardly affected at all
in the range of $q$ considered here. Therefore, as $q$ is decreased,
the finite-size scaling line moves below the HRT points.
Also in this case we found good overall agreement for each $q$.
In the same figures
we have plotted also the HRT results for the correlation length $\xi$ which
governs the decay of the correlations near the critical point.
The divergence of $\xi$ on the critical isochore asymptotically close
to the critical point is described by 
the power law $\xi\sim C|t|^{-\nu}$. The current
implementation of HRT necessarily gives a vanishing critical exponent $\eta$.
Therefore, the critical exponent relation between $\gamma$
and $\nu$ becomes just $\nu=\gamma/2$.
As a general remark, we observe that, while the FSS curves represent
the asymptotic power-law behavior and therefore give a straight line
on the log-log plots of \fig{FIG2}-\fig{FIG7}, the HRT results do not,
since in the crossover region the corrections to the asymptotic scaling
are important. These corrections are nonuniversal, and depend on the size
ratio $q$.

It is worthwhile mentioning that the critical
behavior of a colloid-polymer mixture of sterically stabilized silica spheres 
and polydimethylsiloxane dispersed in cyclohexane  
has been investigated experimentally 
\cite{chen.payandeh:2000, chen.payandeh:2001}. 
The isothermal compressibility, the correlation 
length \cite{chen.payandeh:2000}, the order parameter, 
and the interfacial tension \cite{chen.payandeh:2001} of the colloid 
were measured as a function of the polymer concentration, and the critical
exponents $\gamma$, $\beta$, $\nu$ were found to be slightly higher than the
asymptotic Ising values. This was explained in terms 
of exponent renormalization 
by the factor $1/(1-\alpha)$, $\alpha$ being the critical exponent 
of the specific heath \cite{fisher}. On the other hand, exponent 
renormalization is not found here, where the quantity driving criticality
is the {\em reservoir} density of the polymer, or equivalently its chemical
potential. This is a field analogous to temperature in thermal systems, 
and does not cause renormalization. 

In summary, we have carried out a study of the critical behavior of a simple
model of colloid-polymer mixtures by both MC simulations and HRT.  
Critical fluctuations in colloidal
systems are expected to have a strong effect
on phase separation,
e.g. lowering the free energy barrier for nucleation of the solid 
in the fluid phase \cite{frenkel}.
More generally, there exist many examples which show the importance of
a full understanding of 
the universality class of the systems in exam.
In a previous paper \cite{vink.horbach.ea:2004}, one of us studied 
the interfacial tension of the AO model, which gives indication 
of the strength of capillary waves.
The mean-field like behavior and the 3D Ising behavior
of the capillarity strength are profoundly different; then   
it is important
to reproduce correctly the universality class of the systems in exam.
The importance of an accurate theoretical approach for this 
study, where the ability of HRT to describe realistically criticality can 
be very helpful, origins also from the numerical efforts and time cost
necessary to the simulation of a full colloid-polymer mixture.

\vspace{10 mm}

\begin{center}
{\bf Acknowledgments}
This work is funded in part by the Marie Curie program of the
European Union, contract number MRTN-CT2003-504712. Support from the {\it
Deutsche Forschungsgemeinschaft} under the SFB-TR6 (project sections A5 and D3)
is also acknowledged. We thank K. Binder and J. Horbach for many stimulating
discussions.

\end{center}

\clearpage

\end{document}